\documentclass[a4paper,12pt]{article}
\usepackage[utf8x]{inputenc}
\usepackage{graphicx}
\usepackage{overcite}
\usepackage{amsmath}
\usepackage{amssymb}
\usepackage{here}

\newcommand{\onlinecite}[1]{\hspace{-1 ex} \nocite{#1}\citenum{#1}}

\title{Heuristics-Guided Exploration of Reaction Mechanisms}
\author{Maike Bergeler, Gregor N.\ Simm, Jonny Proppe, and Markus Reiher\thanks{corresponding author: markus.reiher@phys.chem.ethz.ch; Phone: +41446334308; Fax: +41446331594}}

\begin{document}

\maketitle

\vspace*{-0.9cm}\begin{center}
ETH Z\"urich, Laboratory of Physical Chemistry, \\ Vladimir-Prelog-Weg 2, 8093 Z\"urich, Switzerland. \\
\end{center}

\begin{abstract}
For the investigation of chemical reaction networks, the efficient and accurate determination of all relevant intermediates and elementary reactions is mandatory.
The complexity of such a network may grow rapidly, in particular if reactive species are involved that might cause a myriad of side reactions.
Without automation, a complete investigation of complex reaction mechanisms is tedious and possibly unfeasible.
Therefore, only the expected dominant reaction paths of a chemical reaction network (e.g., a catalytic cycle or an enzymatic cascade) are usually explored in practice.
Here, we present a computational protocol that constructs such networks in a parallelized and automated manner.
Molecular structures of reactive complexes are generated based on heuristic rules derived from conceptual electronic-structure theory 
and subsequently optimized by quantum chemical methods to produce stable intermediates of an emerging reaction network.
Pairs of intermediates in this network that might be related by an elementary reaction according to some structural similarity measure
are then automatically detected and subjected to an automated search for the connecting transition state.
The results are visualized as an automatically generated network graph, from which a comprehensive picture of the mechanism of a complex chemical 
process can be obtained that greatly facilitates the analysis of the whole network. 
We apply our protocol to the Schrock dinitrogen-fixation catalyst to study alternative pathways of catalytic ammonia production.
\end{abstract}

\section{Introduction}
Complex reaction mechanisms are found in transition-metal catalysis \cite{Masters1980}, polymerizations \cite{Vinu2012}, cell metabolism \cite{Ross2008}, flames and environmental processes \cite{Vereecken2015}, and are the objective of systems chemistry \cite{Ludlow2007}.
Knowing all chemical compounds and elementary reactions of a specific chemical process is essential for its understanding in atomistic detail.
Even though many chemical reactions result in the selective generation of a main product \cite{Clayden2001}, generally, multiple reaction paths compete with each other leading to a variety of side products.
In such cases, a reactive species (such as a radical, a valence-unsaturated species, a charged particle, a strong acid or base) can be involved or the energy deposited into the system may be high (e.g., due to a high reaction temperature). 

For a detailed analysis of a chemical system, relevant intermediates and transition states are to be identified according to their relative energies.
Manual explorations of complex reaction mechanisms employing well-established electronic-structure methods are slow, tedious, and error-prone.
They are therefore limited to the search for expected dominant reaction paths. 
It is therefore desirable to develop a fully automated protocol for an efficient and accurate exploration of configuration spaces involving both intermediates and transition states.

Existing approaches comprise, for example, global reaction route mapping \cite{Maeda2013}, and reactive molecular dynamics \cite{vanDuin2001,Dontgen2015,Saitta2014,Wang2014}.
Starting from a given structure, the global-reaction-route-mapping procedures evolves along the corresponding potential-energy surface (PES) by exploiting local curvature information.
Since the dimension of a PES scales with the number of atomic nuclei, the global-reaction-route-mapping methods, though highly systematic, are not suitable for the exploration of large reactive systems.
Contrary to such stationary approaches, in reactive-molecular-dynamics simulations, the nuclear equations of motion are solved to explore and sample configuration spaces.
Recently, the capability of reactive ab initio molecular dynamics for studying complex chemical reactions was shown at the example of the prebiotic Urey--Miller experiment \cite{Saitta2014,Wang2014}.
To overcome the high computational demands of first-principles calculations in ab initio molecular dynamics, 
reactive force fields \cite{vanDuin2001} can be employed, which accelerate calculations by some orders of magnitude \cite{Dontgen2015}.
They are, however, not generally available for any type of system.

A complementary strategy is to exploit the conceptual knowledge of chemistry from quantum mechanics to explore reaction mechanisms \cite{Earis2007}.
By applying predefined transformation rules to create new chemical species based on reactivity concepts, searches in the chemical configuration space are accelerated without resorting to expert systems as applied for synthesis planning \cite{Corey1976,Pensak1977,Gasteiger1990,Rucker2004,Todd2005,Grzybowski2009,Chen2009,Fuller2012,Kowalik2012}.
In this context, the efficiency of heuristics-guided quantum chemistry was explored very recently \cite{Zimmerman2013, Zimmerman2015a, Rappoport2014, Zubarev2015}.
The idea behind heuristic guidance in quantum chemistry is to propose a large number of hypothetical molecular structures, which are subsequently optimized by electronic-structure methods.

Here, we describe a fully automated and parallelized exploration protocol in which the synergies of chemical concepts and electronic-structure methods are exploited.
We implement a heuristics-guided setup of reactive complexes by placing a reactive species in the vicinity of a target reactive site so that a 
high-energy structure emerges.
Hence, we construct 'reactive complexes' as supermolecular structures of reactants that are sufficiently high up the Born--Oppenheimer potential energy 
surface to produce a reaction product upon structure optimization. The reactive complexes are then optimized with quantum-chemical methods 
to eventually yield stationary points in the vicinity of 
these high-energy structures. The stationary points finally enter as vertices an automatically generated reaction network.
Subsequently, elementary reactions between pairs of stationary points are identified and validated by automatically launched transition-state searches and intrinsic-reaction-path calculations, respectively.
Finally, the results are visualized as automatically generated network graphs endowed with thermodynamic and kinetic parameters.

To illustrate the functionality of our protocol, we apply it to an important and still not well understood problem in chemistry, that is
catalytic nitrogen fixation under ambient conditions in the homogeneous phase.
For this purpose, we investigate the chemical reactivity of the molybdenum complex developed by Schrock and co-workers \cite{Yandulov2002,Yandulov2003,Schrock2008}.
This catalyst, like all others developed for this purpose \cite{Hinrichsen2012,Arashiba2011,Lee2010}, is plagued by a very low turnover number.
By applying our protocol to a simplified model system, we aim to better understand the low efficiency of the catalyst.

\section{Heuristic Guidance for Quantum-Chemical Structure Explorations}
In the context of reaction mechanisms, heuristic rules serve to propose the constituents of a chemical reaction network, which, when optimized, will be the minimum-energy structures and transition states that are energetically accessible at a given temperature.
Although this approach cannot guarantee to establish a complete resulting reaction network, heuristic methods allow for a highly efficient and directed search based on empiricism and chemical concepts.

Crucial for the construction of such heuristic rules is the choice of molecular descriptors.
For the study of chemical reactions, graph-based descriptors dominate the field \cite{Rucker2004, Todd2005, Chen2008, Chen2009, Graulich2010, Kayala2011, Kayala2012, Zimmerman2013,Zimmerman2015a, Rappoport2014, Zubarev2015}, which are based on the concept of the chemical bond.
Zimmerman \cite{Zimmerman2013, Zimmerman2015a} developed a set of rules based on the connectivity of atoms to generate molecular structures and to determine elementary reactions. 
Quantum-chemical structure optimizations and a growing-string transition-state-search method \cite{Zimmerman2013a, Zimmerman2013b, Zimmerman2015} were applied to study several textbook reactions in organic chemistry. 
Aspuru-Guzik and co-workers \cite{Rappoport2014, Zubarev2015} developed a methodology for testing hypotheses in prebiotic chemistry. Rules based on formal bond orders and heuristic functions inspired by Hammond's postulate to estimate activation barriers were applied to model prebiotic scenarios and to determine their uncertainty.
Very recently, a new algorithm for the discovery of elementary-reaction steps was published \cite{Suleimanov2015} that uses freezing-string and Berny-optimization methods to explore new reaction pathways of organic single-molecule systems.
While graph-based descriptors perform well for many organic molecules, they may fail for transition-metal complexes, where the chemical bond is not always well defined \cite{Frenking2014}.

\begin{figure}[H]
\begin{center}
\includegraphics[width=0.45\textwidth]{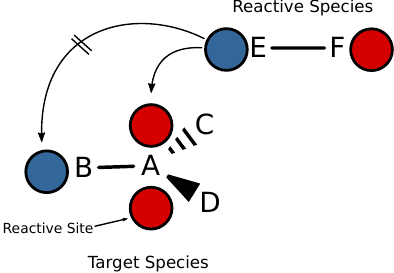}
\caption{A reactive site of a reactive species approaching the reactive sites of a target species.
The color of a reactive site represents the value of some chemical-reactivity descriptor.}
\label{fig:reactive_site}
\end{center}
\end{figure}

Complementary to Zimmerman's and Aspuru-Guzik's approaches, we aim at a less context-driven method to be applied to an example of transition-metal catalysis.
Clearly, such an approach must be based on information directly extracted from the electronic wave function so that no 
additional (ad hoc) assumptions on a particular class of molecules are required.
In the first step of our heuristics-guided approach, we identify \textit{reactive sites} in the chemical system.
When two reactive sites are brought into close proximity, a chemical bond between the respective atoms is likely to be formed
(possibly after slight activation through structural distortion).
In addition, we define \textit{reactive species} which can attack \textit{target species} at their reactive sites.
This concept is illustrated in Fig.~\ref{fig:reactive_site}.

A simple example for the first-principles identification of reactive sites is the localization of Lewis-base centers in 
a molecule as attractors for a Lewis acid. 
Lone pairs are an example for such Lewis-base centers and can be detected by inspection of an electron localization measure such
as the electron localization function (ELF) by Becke and Edgecombe \cite{Becke1990} or the Laplacian of the electron density 
as a measure of charge concentration \cite{Bader1994} (see also Ref.\ \onlinecite{Ayers2005}). 
Other quantum chemical reactivity indices can also be employed\cite{Earis2007}, such as 
Fukui functions \cite{Fukui1982}, partial atomic charges \cite{Mulliken1955, Mulliken1955a, Lowdin1970}, or atomic 
polarizabilities \cite{Brown1982, Kang1982}.
With these descriptors, reactive sites can be discriminated, i.e., not every reactive site may be a candidate for every 
reactive species (indicated by the coloring in Fig.~\ref{fig:reactive_site}).
For example, an electron-poor site is more likely to react with a nucleophile rather than with an electrophile.
Moreover, reactive species consisting of more than one atom may have distinct reactive sites.
Naturally, the spatial orientation of a reactive species toward a reactive site is important.

In the second step, reactive species are added to a target species resulting in a set of candidate structures for reactive complexes.
Such compound structures should resemble reactive complexes of high energy (introduced by sufficiently tight structural positioning of the reactants,
optionally activated by additional elongation of bonds in the vicinity of reactive sites) which are then optimized employing 
electronic-structure methods (third step). By means of standard structure-optimization techniques we search for potential reaction products for the
reaction network from the high-energy reactive complexes.
Several structure optimizations of distinct candidates may result in the same minimum-energy structure. 
Such duplicate structures must be identified and discarded to ensure the uniqueness of intermediates in the network (fourth step).
It should be noted that each intermediate of a reaction network can be considered a reactive species to every other intermediate of that network.

Through a structural comparison [based on a distance criterion such as the root-mean-square deviation (RMSD)], 
pairs of structures which can be interconverted by an elementary reaction, i.e., a single transition state, are identified (fifth step).
If no such pair can be found for a certain structure, the local configuration space in the vicinity of that structure needs to be explored further to ensure that no intermediate will be overlooked.

In the sixth step, the automatically identified elementary reactions are validated by transition-state searches and subsequent intrinsic-reaction-path calculations.
Several automated search methods for transition states are available such as interpolation methods \cite{Halgren1977}, eigenvector following \cite{Cerjan1981, Simons1983, Wales1992, Wales1993, Jensen1995}, string methods \cite{Henkelman2000, Behn2011, Zimmerman2013a}, the scaled-hypersphere-search method \cite{Ohno2004}, constrained optimization techniques \cite{Gonzalez1990}, quasi-Newton methods \cite{Broyden1967, Fletcher1980}, Lanczos-subspace-iteration methods \cite{Malek2000} and related techniques \cite{Henkelman1999,Munro1999}, 
or Davidson-subspace-iteration-based algorithms \cite{Bergeler2015}).

In the seventh and last step of our heuristics-guided approach, a chemical reaction network comprising all determined intermediates and transition states is automatically generated.
The visualization of results as network graphs in which vertices and edges represent molecular structures and elementary reactions, respectively, supports understanding a chemical process in atomistic detail.
The readability of a network graph can be enhanced if vertices and edges are supplemented by attributes such as colors or shapes 
chosen with respect to their relative energy or to other physical properties.

Even though our heuristics-guided approach aims at restricting the number of possible minimum-energy structures, the number of generated intermediates may still be exhaustively large as the following example illustrates.
For a protonation reaction, we may assume that the number of different protonated intermediates can be determined from the unprotonated target species by identifying all reactive sites (RS) which a proton, the reactive species, can attack.
This number is given by a sum of binomial coefficients,
\begin{equation}
N = \sum_{p=1}^{n_{\mathrm{RS}}} {n_{\mathrm{RS}} \choose p}=2^{n_{\mathrm{RS}}}-1 \label{eq:np},
\end{equation}
where $n_{\mathrm{RS}}$ is the number of reactive sites and $p$ is the number of protons added to the target species. 
Even for such a simple example, the number of possible intermediates increases exponentially. 
For example, for a target species with ten reactive sites, $N = 1023$ intermediates will be generated.
Obviously, the transfer of several protons to a single target species is not very likely from a physical point of view as charge will increase so that the acidity of the protonated species might not allow for further protonation.
In the presence of a reductant, however, these species can become accessible in reduced form.

\section{Construction of Complex Reaction Networks}
Of all chemical species generated by the application of heuristic rules, some will be kinetically inaccessible under certain physical conditions.
By defining reaction conditions (in general, a temperature $T$) and a characteristic time scale of the reaction under consideration, 
one can identify those species that are not important for the evolution of the reactive system under these conditions.
Even if these intermediates are thermodynamically favored, they may not be populated on the characteristic time scale at temperature $T$.
By removing these species from the network, one can largely reduce its complexity, which in turn simplifies subsequent analyses (such as kinetics simulations 
as, for instance, presented in Ref.\ \onlinecite{Glowacki2012}).

For the following discussion, we introduce the notation that a  \textit{chemical reaction network} (or \textit{network}) is to be understood as a connected graph built from a set of intermediates (vertices) and a set of elementary reactions (edges).
A \textit{path} shall denote a directed sequence of alternating vertices and edges, both of which occur only once.
A \textit{subnetwork} is a connected subgraph of a network uniquely representing a single PES defined by the number and type of atomic nuclei, the number of electrons, and by the electronic spin state. 
Subnetworks can be related to each other according to the heuristic rules which describe addition or removal of reactive species (defined by their nuclear framework, i.e., by their nuclear attraction potential and charge) and electrons.
The initial structures are referred to as \textit{zeroth-generation structures}, and generated structures are referred to as \textit{higher-generation structures}.
\textit{Substrates} are species that represent the reactants of a complex chemical reaction.
The initial population of all other target species is zero.

For the exclusion of non-substrate vertices from the reaction network, we propose a generic energy-cutoff rule:
\textit{If each path from a substrate vertex to a non-substrate vertex comprises at least one sequence of consecutive vertices with an 
increase in energy larger than a cutoff $E_\text{C}$, then we remove the non-substrate vertex from the network.}

\begin{figure}[H]
\begin{center}
\includegraphics[width=\textwidth]{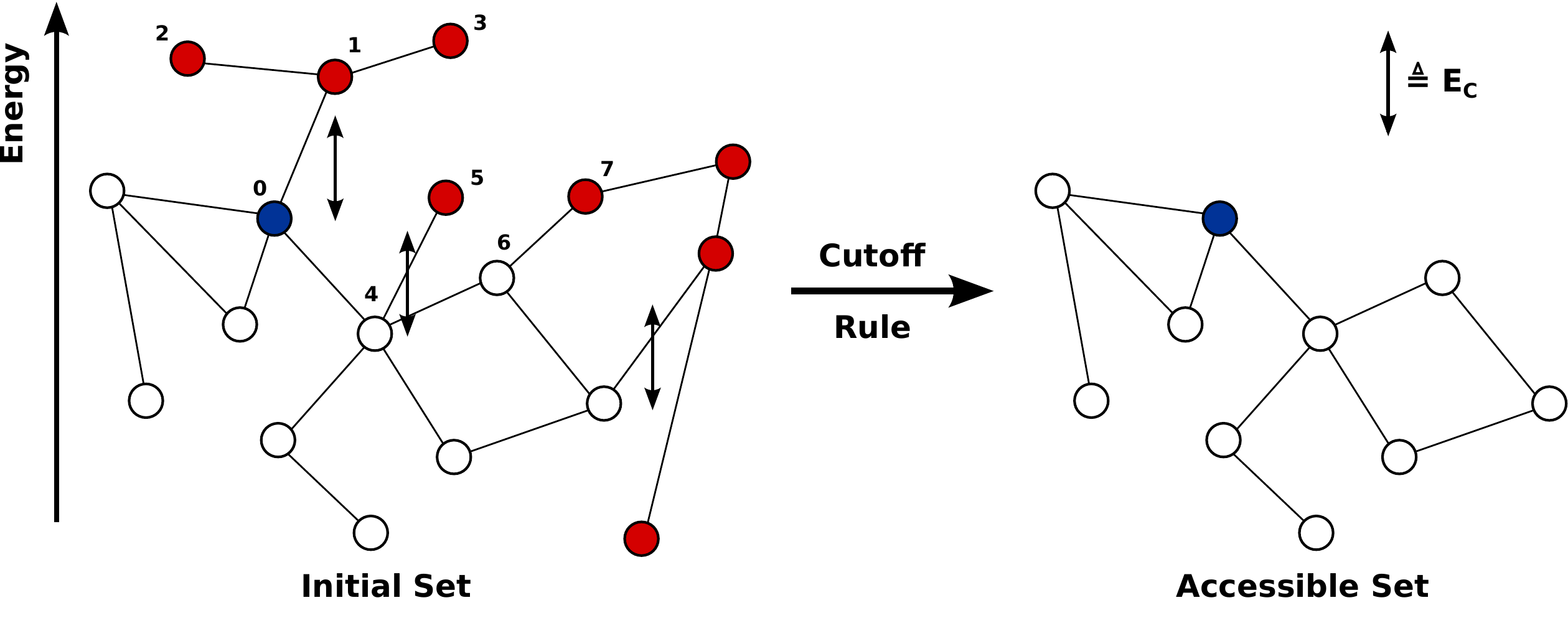}
\caption{Illustration of the process of removing intermediates (shown as vertices) from a chemical reaction network by applying the energy-cutoff $E_\text{C}$.
  The vertex representing the substrate is blue-colored, vertices to be removed are red-colored.
}
\label{bubbles}
\end{center}
\end{figure}
The application of this energy-cutoff rule is illustrated in Fig.~\ref{bubbles}.
Starting from substrate 0, intermediate 1 can only be reached via a transition-state higher than $E_\text{C}$, and therefore, it can be removed from the network.
Since intermediates 2 and 3 can only be reached via intermediate 1, they can also be omitted. 
Despite being similar in energy to substrate 0, intermediate 5 can be discarded, since it can only be formed by a transition state higher than $E_\text{C}$.
Even though the transition state between intermediates 6 and 7 is below $E_\text{C}$, the population of intermediate 7 is negligible, since, starting from substrate 0, it can only be formed through intermediate 4.

Note that this energy-cutoff rule is conservative as we compare energy differences of stable intermediates, which are a lower bound for 
activation energies of reactions from a low-energy intermediate to one that is higher in energy.
Therefore, intermediates can be removed prior to the calculation of transition-state structures, which significantly saves computational resources.
Once transition states are calculated, this rule can be reapplied to further reduce the complexity of the network in order to arrive at a minimal network of all relevant reaction steps.

The introduced kinetic cutoff $E_\text{C}$ depends on temperature $T$ and on the characteristic time scale of the reaction.
For instance, assuming a reactive system following Eyring's quasi-equilibrium argument \cite{Eyring1935}, one can determine the average time for a unimolecular reaction to occur.
For this purpose, the general Eyring equation,
    \begin{equation}
     k=\frac{k_\text{B}T}{h}\exp\Bigg(-{\Delta G^{\ddagger}\over{RT}}\Bigg),
     \label{eyring}
     \end{equation}
     is employed, with the rate constant $k$, the Boltzmann constant $k_\text{B}$, the Planck constant $h$, the Gibbs free energy of activation $\Delta G^{\ddagger}$, and the temperature $T$.
     We understand the half life $\text{ln}(2) / k$ as the time after which a molecule has reacted with a probability of 50\%.
     For an activation free energy of 25~kcal/mol and a temperature of $T = 298$ K, the average time for a unimolecular reaction to occur equals three days. 
     This time may well be considered an upper limit for a practical chemical reaction.
     If one can afford longer reaction times, the energy cutoff needs to be increased.
     Similarly, if one is interested in a range of temperatures, $\Delta T = T_\text{max}-T_\text{min}$, the energy cutoff has to be adapted to the maximum temperature $T_\text{max}$.
     Otherwise, intermediates would be removed from the network which are accessible at $T_\text{max}$.
     In a conservative exploration, a reasonable choice for the maximum temperature may be the decomposition temperature of an important compound class studied.

     Special attention needs to be paid to the energy differences between intermediates of different subnetworks, since our protocol divides the PES of the chemical system into various subsystem PES's.
     For instance, if two intermediate structures differ by one reactive species, say a proton, the energy for supplying that reactive species by a strong acid has to be taken into account.
     Otherwise, different subnetworks of a network cannot be compared as the total energies to be compared depend on the number of (elementary) particles.

\section{The Chatt--Schrock Network}

\begin{figure}
\begin{center}
\includegraphics[width=\textwidth]{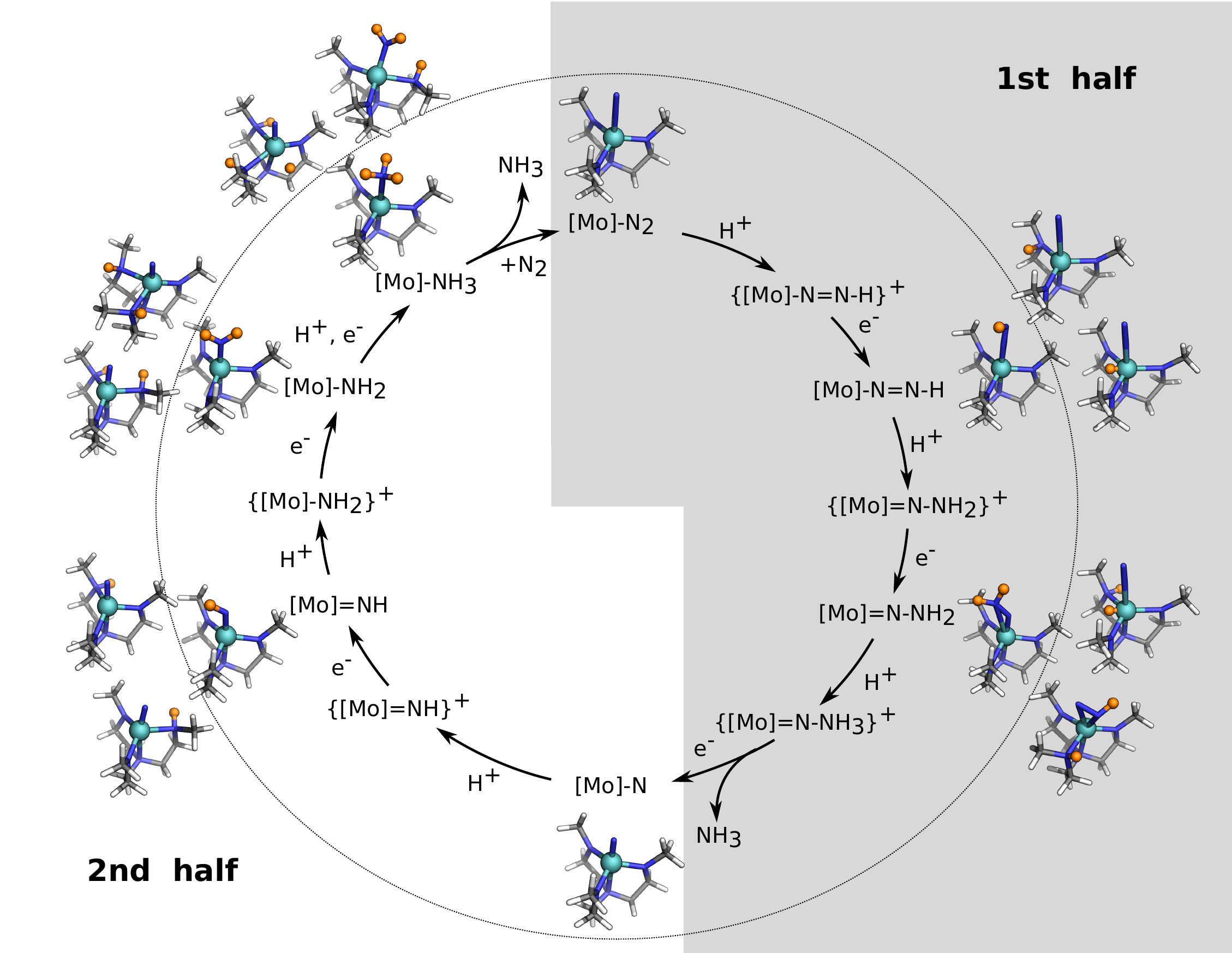}
\caption{The Chatt--Schrock nitrogen-fixation cycle:
The first and second half of the catalytic cycle are based on the [Mo]--N$_2$ and [Mo]--N scaffolds, respectively.
Molecular structures within the circle represent Schrock intermediates.
A selection of isomers of these Schrock intermediates is shown outside the circle.
Element color code: gray, C; blue, N; turquoise, Mo; white, H; orange, H added to reactive sites.}
\label{fig:cycle}
\end{center}
\end{figure}

The (generic) Chatt--Schrock cycle \cite{Yandulov2002,Yandulov2003} (Fig.~\ref{fig:cycle}) is the prominent example of catalytic nitrogen fixation \cite{Hinrichsen2012}.
Its intermediates (referred to as \textit{Schrock intermediates} hereafter) are formed by an alternating sequence of single protonation and single electron-reduction steps of Schrock's nitrogen-ligated molybdenum complex \cite{Yandulov2002, Yandulov2003}.
The sources of protons and electrons are 2,6-lutidinium (2,6-LutH) and decamethylchromocene (Cr$\text{Cp}^{\ast}_2$), respectively.
This mechanism, however, does not explain the small turnover number of the catalyst.
To demonstrate our heuristic network-exploration algorithm described above, we aim at identifying competing reaction paths of the Chatt--Schrock cycle.
For details on the Computational Methodology, see the appendix.

\subsection{Heuristics-Guided Structure Search}
\begin{figure} [H]
\begin{center}
\includegraphics[width=0.5\textwidth]{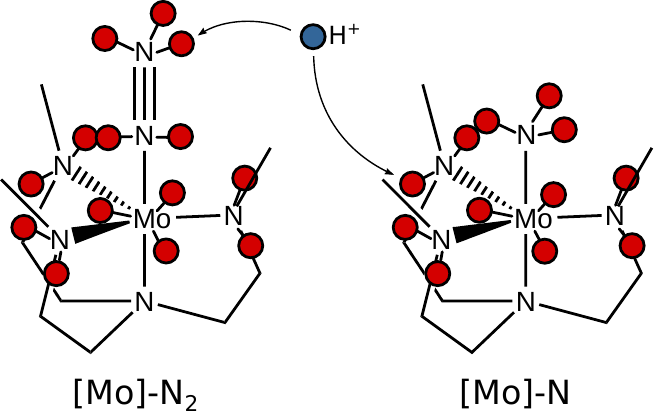}
\caption{The reactive species (H$^+$) attacking the reactive sites (proton-accepting red circles) of [Mo]-N$_2$ and [Mo]-N. 
Lines are drawn between an atom and its reactive sites to highlight their spatial arrangement.
Each amide nitrogen atom exposes two protonation sites, although we occupy at most one to produce an amine nitrogen atom.
}
\label{fig:sites}
\end{center}
\end{figure}

For the first and second half of the catalytic cycle (Fig.~\ref{fig:cycle}) [Mo]--N$_2$ and [Mo]--N (Fig.~\ref{fig:sites}) are taken as zeroth-generation structures, respectively.
Here, [Mo] refers to the Yandulov--Schrock complex \cite{Yandulov2002,Yandulov2003} where the hexa-\textit{iso}-propyl terphenyl (HIPT) substituents are replaced by methyl groups to reduce the computational cost.
The bulky HIPT substituents can be reintroduced once the network has been established.

In this study, we only consider protons as reactive species since protonations of the amide nitrogen atoms are likely to deactivate the catalyst \cite{Schrock2008,Schenk2008}.
Additionally, we take different charges of the protonated complexes into account (single electron-reduction steps from $y$+ to neutral, with $y$ being the number of protons added).
However, for a more extensive exploration, H$_2$, N$_2$, NH$_3$, N$_x$H$_y$, and intermediates theirselves must also be considered as reactive species.

\begin{figure}[h!]
\begin{center}
 \includegraphics[width=0.7\textwidth]{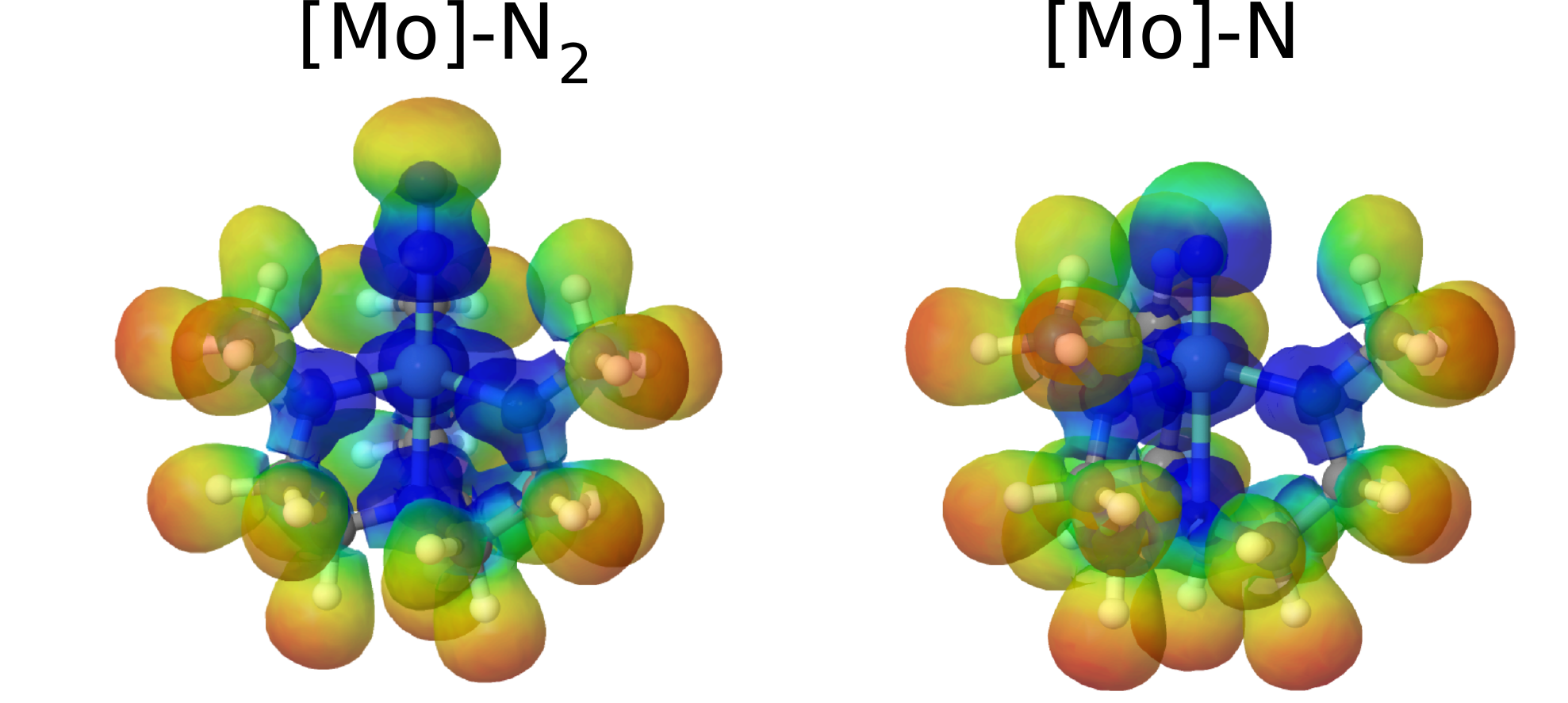}
 \caption{Electron localization function (ELF) for [Mo]-N$_2$ (left) and [Mo]-N (right) colored according to the electrostatic potential (an 
isosurface value for ELF of 0.6 a.u. was chosen).}
\label{elf}
\end{center}
\end{figure}

To determine the reactive sites of the substrates, we exploit knowledge about negative charge concentrations extracted from
the electronic wave function. As an example in Fig.\ \ref{elf}, we present the isosurface of the ELF colored with the value of the
electrostatic potential for the two parent species, [Mo]-N$_2$ and [Mo]-N, of the two halves of the Chatt--Schrock cycle in Fig.\ \ref{fig:cycle}.
Whereas the ELF highlights regions in space where electron density is localized, the electrostatic potential allows us to pick those
regions that can function as a Lewis base (highlighted in blue in Fig.\ \ref{elf} and showing, e.g.,  lone pairs)
by contrast to the other regions that are electron deficient and feature hardly any Lewis basicity 
(highlighted in orange in Fig.\ \ref{elf} and showing, e.g., C--H $\sigma$-bonds).
The blue regions therefore define spatial areas that function as reactive sites to which protons should be added as reactive species.

Note that this procedure is solely based on the first principles of quantum mechanics, but that it is also in line
with conventional chemical wisdom that, for
carbon and nitrogen atoms, the formation of a tetrahedral surrounding including both bonding neighbors and reactive sites for incoming reactants (protons)
represents a valence-saturated electronic situation.
At the molybdenum center, three reactive sites are introduced in the plane spanned by the three amide-nitrogen atoms.
We refrain from protonating the coordinating amine nitrogen atom in {\it trans}-position to the N$_2$ ligand as this
would produce a decomposition pathway of the catalyst that is not likely to lead to an alternative catalytic cycle. Clearly, these
decomposition reactions are important to track for a complete understanding of Schrock-type dinitrogen fixation catalysis, but we
devote this aspect to future work. Instead, we are less restrictive with respect to the possible protonation sites at the metal center and at the terminal
nitrogen atom of N$_2$ in [Mo]-N$_2$ --- density-functional-theory calculations are fast for the size of system under
study and can be carried out in parallel so that one should not limit the number of possible reactive sites too much in order
not to risk overlooking of important intermediates.
All reactive sites considered as proton-acceptor sites in this work are shown in Fig.~\ref{fig:sites}.
Up to four protons are added combinatorially to the zeroth-generation structures.
This number may be considered a chemically reasonable upper limit.

Since the Yandulov--Schrock catalyst operates in the presence of a strong reducing agent (Cr$\text{Cp}^{\ast}_2$), protonated species can be readily reduced.
Therefore, for a $p$-fold protonated reactive complex, we consider the charges $0\leq C\leq p$.
The result is a total number of
\begin{equation}
N=\sum_{p=1}^{n_\text{RS}} \left\{ {n_\text{RS} \choose p} (p+1) \right\}
\end{equation}
structures.
For [Mo]-N$_2$ $N$=6762 and for [Mo]-N $N$=3577 structures are obtained.
However, for the subsequent structure optimizations these numbers are slightly reduced as the two protonation sites exposed by each amide nitrogen atom are
occupied by at most one proton to yield an amine nitrogen atom.

Decomposed reactive complexes such as those from which dihydrogen dissociated or in which the chelating ligand (partially) dissociated from the metal center, are automatically removed from the network, but stored for future investigation (see also Supporting Information).

To ensure the uniqueness of each vertex in the network, duplicate structures are removed.
For this, the threefold symmetry of the catalyst was taken into account. 

Subsequently, elementary reactions connecting two vertices are identified.
We determine two intermediates of the same subnetwork to be constituents of an elementary reaction if they are related by a shift of a single proton, i.e., 
exactly one proton is required to change its position, while the change of the remaining molecular structure is below a predefined RMSD cutoff of 
0.5 {\AA} for the first half and 0.65 {\AA} for the second half of the Chatt--Schrock cycle.
For elementary reactions between subnetworks, the same RMSD cutoff was chosen to determine the shared identity of two molecular structures (apart from an added proton in case of a protonation reaction).
Clearly, also in general case some criterion for structural similarity of two vertices may be used to assess whether they 
are connected by an elementary reaction step.

Subsequently, transition-state searches based on electronic-structure methods are performed for the proposed intra-subnetwork elementary reactions.
We employed constrained optimizations to generate reasonable guess structures, which are refined by an eigenvector-following algorithm (see the Computational Methodology in the appendix for details) and verified by Mode-Tracking calculations \cite{Reiher2003, Bergeler2015}.
Finally, intrinsic-reaction-path calculations are performed.
For intermolecular reactions (i.e., proton transfers from 2,6-LutH), no transition states were calculated.

In this study, we applied an energy cutoff of $E_\text{C}$ = 25 kcal/mol for the relative energy of two stable intermediates connected by an elementary reaction.
Note that this threshold does not refer to a Gibbs free reaction energy but to an electronic-energy difference.
Nevertheless, since only reactions of the same type are compared, one can expect only small deviations of electronic energies from Gibbs free energies for intranetwork reactions (proton shifts) and reduction steps, and assume that this simplification is also a good approximation for protonation reactions.

\subsection{Network Superstructure}
\begin{figure}[H]
\begin{center}
\includegraphics[width=0.7\textwidth]{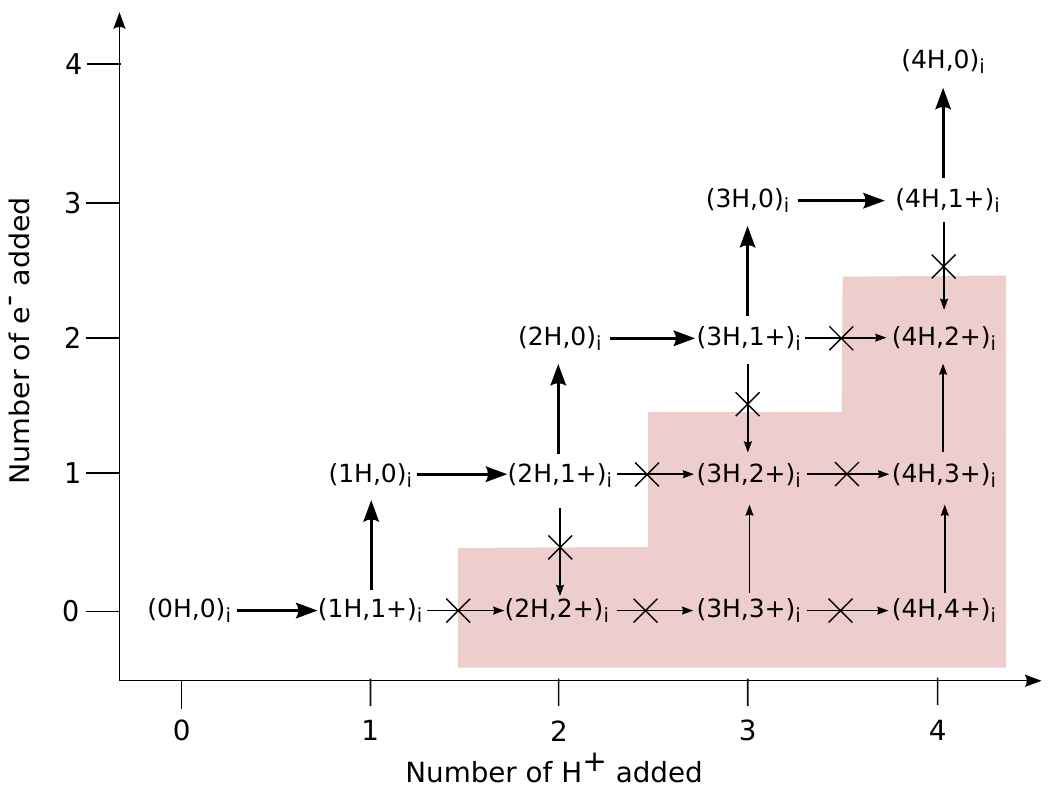}
\caption{
Energy analysis of Schrock's catalytic nitrogen fixation.
Subnetworks in the red shaded area cannot be reached from their respective substrate 
$i$ (1 = [Mo]-N$_2$, 2 = [Mo]-N) without exceeding a transition-state energy above $E_\text{C}$.
}
\label{fig:deltaEa}
\end{center}
\end{figure}

In Fig.~\ref{fig:deltaEa}, subnetworks are arranged according to the number of protons and electrons added.
Here, the subnetworks are denoted as ($x$H, $c$)$_i$, where $x$ is the number of hydrogen atoms added to the substrate $i$ (1 = [Mo]-N$_2$, 2 = [Mo]-N) and $c$ is the charge of the subnetwork.
An arrow pointing from subnetwork \textit{a} to subnetwork \textit{b} will be crossed out if all intermediates of subnetwork \textit{b} are at least by $E_\text{C}$ energetically higher than all intermediates of subnetwork \textit{a}.
In such cases, our cutoff rule can be applied to entire subnetworks since the lowest possible transition barrier from subnetwork \textit{a} to subnetwork \textit{b} will be larger than $E_\text{C}$.

Due to the energy-cutoff rule, entire subnetworks can be pruned and excluded from further analysis.
For instance, starting from (0H,0)$_1$, (2H,2+)$_1$ cannot be reached via any other subnetwork without having to overcome a transition state that is above $E_\text{C}$. 
Therefore, (2H,2+)$_1$ can be removed from the network.
In both halves of the Chatt--Schrock cycle, all subnetworks with a total charge larger than one can be neglected.
The pruning of these networks largely reduces the complexity of the network since now every subnetwork can only be reached from one other subnetwork.

\begin{figure}[H]
\begin{center}
\includegraphics[width=0.7\textwidth]{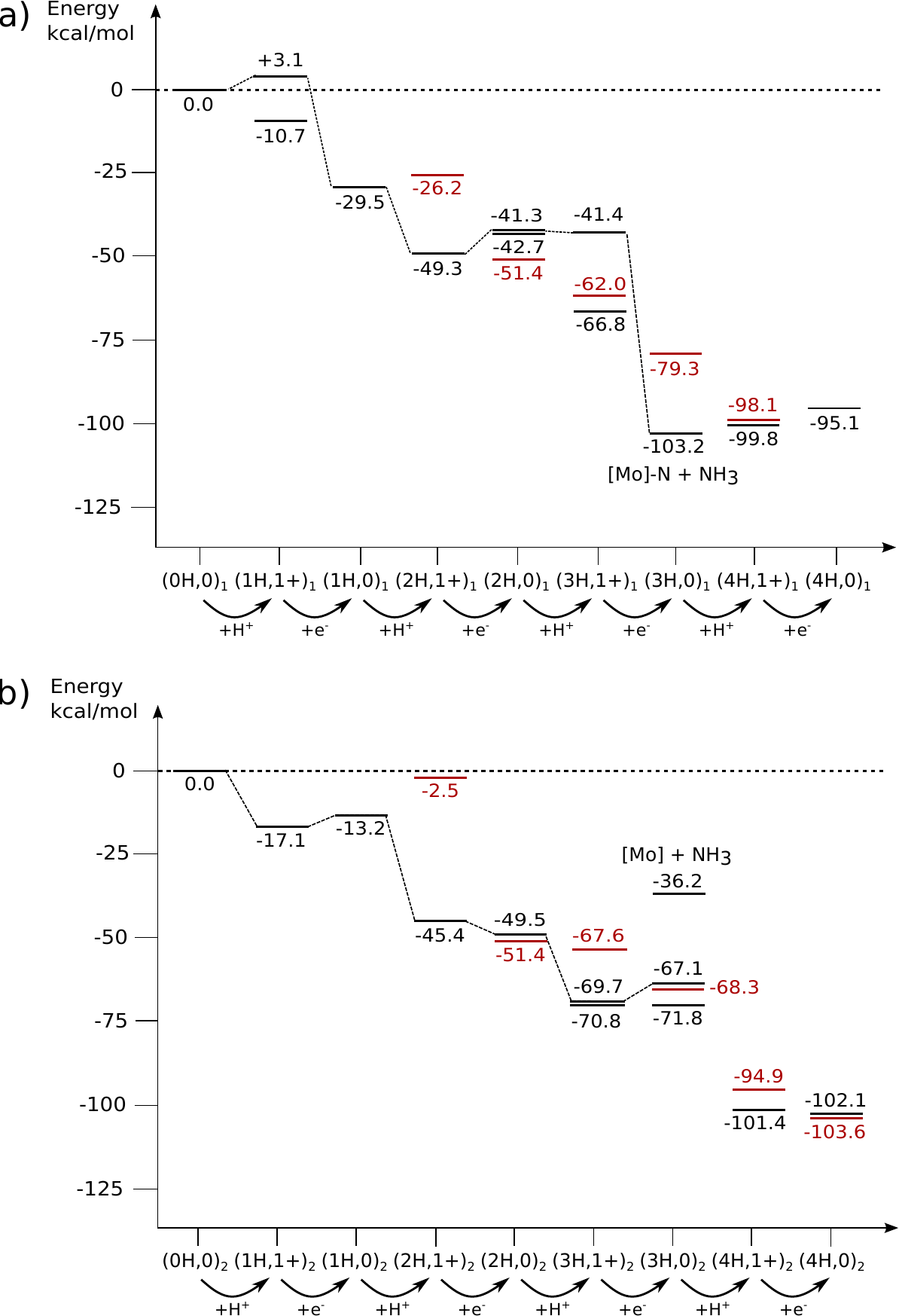}
\caption{
Energy profile of the Chatt--Schrock cycle [first half: a); second half: b)] including intermediates lower in energy than the Schrock intermediates 
of the inner circle in Fig.\ \ref{fig:cycle}.
Schrock intermediates are connected by dashed lines.
Intermediates from which H$_2$ dissociated are given in red.
The oxidation potential of Cr$\text{Cp}^{\ast}_2$ and the dissociation energy of LutH were calculated (BP86/def2-SV(P)) to be $+103.7$ and $-237.7$ kcal/mol, respectively.
For further details see Supporting Information.
}
\label{fig:deltaEb}
\end{center}
\end{figure}

The energy profiles of the first and second half of the catalytic cycle are shown in Fig.~\ref{fig:deltaEb} a) and b), respectively.
In this figure, energy levels of Schrock intermediates are connected by dashed lines.
An additional energy level will be shown if an intermediate lower in energy than the Schrock intermediate is part of that subnetwork.
Moreover, if intermediates from which H$_2$ dissociated are observed, the intermediate with the lowest energy will be shown in red.

It can be seen that most reactions of the first half of the Schrock cycle are exothermic.
Especially the reductions of the Schrock intermediates of (1H,1+)$_1$ and (3H,1+)$_1$ are thermodynamically favorable.
In addition, the dissociation of NH$_3$ from (3H, 0)$_1$ has a highly negative reaction energy. 
Note, however, that this dissociation energy holds for a specific choice of acid and reductant so that 
the assignment of this reaction energy solely to the breaking of the N--N bond would be misleading 
as discussed in our earlier work \cite{Reiher2005,LeGuennic2005}.

There are also endothermic reactions.
For example, the protonation of the Schrock intermediate of (0H,0)$_1$, was calculated to have a positive reaction energy of +3.1~kcal/mol.
A thermodynamically favorable alternative to the protonation of N$_2$ is the protonation of the amide of the chelate ligand. 
This intermediate is lower in energy ($\Delta E = -13.8$~kcal/mol) than the Schrock intermediate.

Also most reactions in the second half are exothermic.
The protonation of the Schrock intermediate in (1H,0)$_2$ is particularly exothermic with a reaction energy of $-$32.2~kcal/mol.
Nonetheless, there are subnetworks in which the Schrock intermediate is not the most stable species.
In (3H,0)$_2$, for instance, there is an intermediate which is more stable ($\Delta E = 4.7$~kcal/mol) than the respective Schrock intermediate.
Furthermore, it can be seen that the dissociation of H$_2$ is thermodynamically favorable in several subnetworks. 
For example, in (2H,0)$_1$ the dissociation of H$_2$ even results in the most stable intermediate. 

However, we should emphasize that those structures which are very similar in terms of energy may be considered equally stable, especially when viewed 
in the light of the quantum chemical methodology chosen here. 
Moreover, one must keep in mind that the network exploration was carried out for a small model complex of the Schrock catalyst with
a double-zeta basis set.

While the dissociation of NH$_3$ is energetically favorable in the first half, it is very unfavorable in the second.
Therefore, a four-coordinate [Mo] intermediate appears unlikely, and an associative exchange mechanism might be favored over the dissociative one as has already been discussed in the literature \cite{LeGuennic2005,Schenk2008,Schenk2009-2}. 
Since further protonation of (3H,0)$_2$ results in low-energy intermediates, we can identify this subnetwork as a possible starting point of degradation.

The results for the Schrock intermediates reported here are in qualitative agreement with those reported earlier by us 
\cite{Reiher2005,LeGuennic2005,Schenk2008,Schenk2009,Schenk2009-2} and by Tuczek and coworkers \cite{Studt2005,Thimm2015}.
Numerical deviations for the Schrock intermediates are mostly due to the choice of a small model structure and the smaller basis set employed in this work.

\subsection{Reaction Network}
\begin{figure}[H]
\begin{center}
\includegraphics[width=0.8\textwidth]{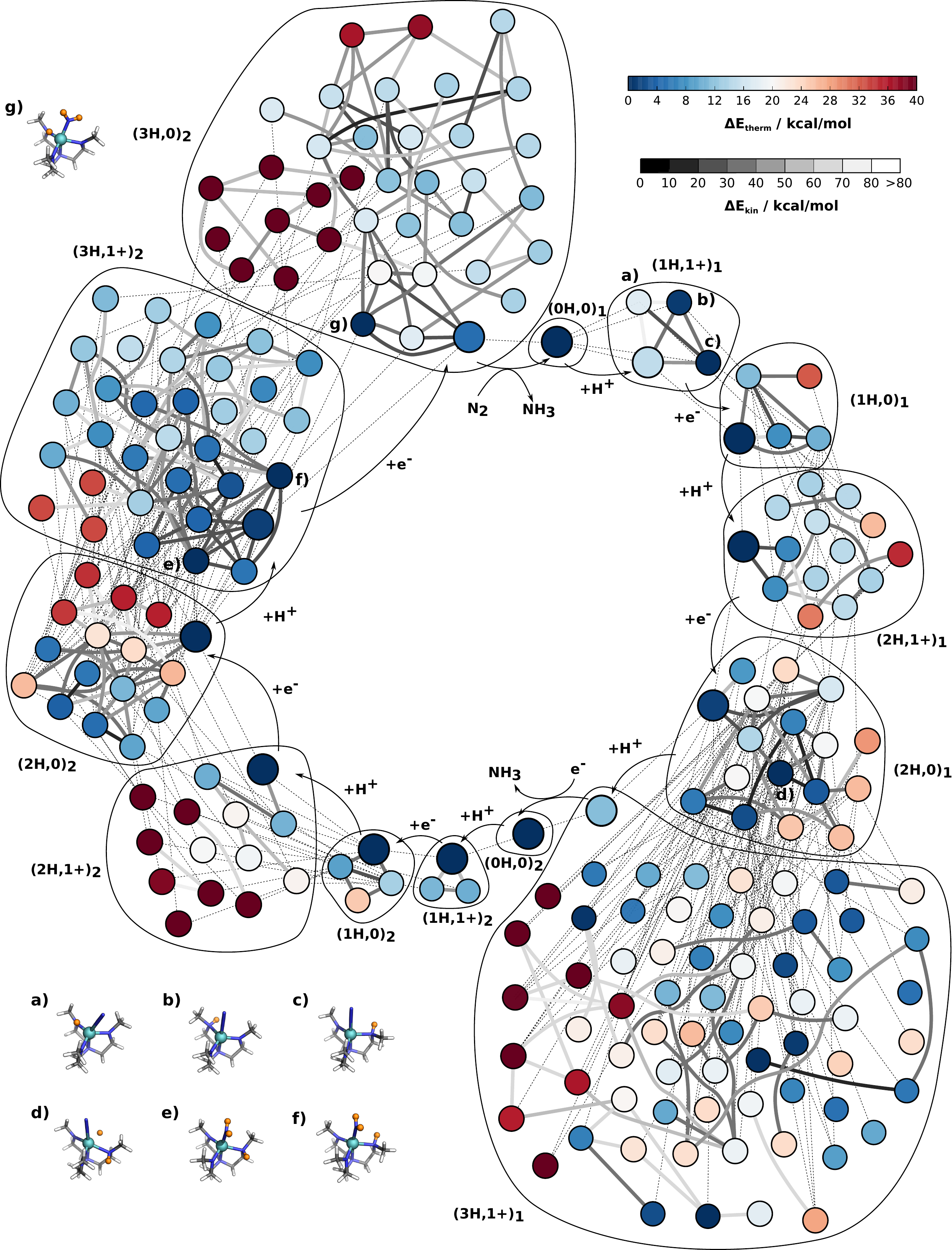}
\caption{
The Chatt--Schrock network of catalytic nitrogen fixation.
Dark-blue vertices refer to the lowest-energy intermediates of a subnetwork, dark-red vertices to the corresponding highest-energy intermediates. 
Vertices representing Schrock intermediates are enlarged. 
Low-energy transition barriers between intermediates of the same subnetwork are indicated by dark-gray edges, high-energy transition barriers by light-gray edges.
Internetwork connections are indicated by dashed lines.
In a)--g) a selection of intermediates is shown.
Element color code: gray, C; blue, N; turquoise, Mo; white, H; orange, H added to reactive sites.
}
\label{fig:network}
\end{center}
\end{figure}

To rationalize the low efficiency and stability of Schrock's nitrogen fixation catalyst, not only all possible intermediates but also the transition states connecting them need to be analyzed.
The reaction network of the Chatt--Schrock cycle automatically generated by our program is shown in Fig.~\ref{fig:network}.
Each vertex represents an intermediate, whereby the color encodes the energy difference with respect to the lowest intermediate of the subnetwork.
Vertices representing a Schrock intermediate are enlarged.
A collection of vertices belonging to the same subnetwork is enclosed by a solid black line.
Two vertices of the same subnetwork are connected by an undirected edge if a transition state was found between them.
The gray scale of such an edge serves as a visual cue indicating the height of the transition barrier with respect to the lower-energy intermediate. Light-gray edges represent high-energy barriers, dark-gray edges represent low-energy barriers.
It is important to note that according to our exclusion rule, many transition states in this network need not to be optimized and can therefore be omitted.
However, to illustrate the complexity of such reaction networks, transition states with an energy above $E_\text{C}$ were not removed.
Vertices of different subnetworks are connected by undirected edges (dashed lines) for which no transition states were calculated.
In addition, the molecular structures of selected intermediates are shown in Fig.~\ref{fig:network} a) -- g).

Starting from the [Mo]-N$_2$ complex in (0H, 0)$_1$, a proton is added to reach (1H,1+)$_1$.
As can be seen from the dashed lines, this reaction can result in four different intermediates. The Schrock intermediate, the [Mo]-N$_2$ complex protonated at the molybdenum atom (Fig.~\ref{fig:network} a)), the [Mo]-N$_2$ complex with a proton at one of the amido groups (Fig.~\ref{fig:network} b)), and the enantiomer of that intermediate (Fig.~\ref{fig:network} c)).
From there, each intermediate can either undergo a reduction to form an intermediate in (1H,0)$_1$, or---through an intramolecular reaction---transform into another intermediate of the same subnetwork.
The subsequent protonation of intermediates in (1H,0)$_1$ leads to the subnetwork (2H,1+)$_1$.

The inspection of the first four subnetworks already suggests a feasible alternative to the Chatt--Schrock mechanism:
The [Mo]-N$_2$ complex is protonated at the amido group; this intermediate undergoes reduction, protonation (of the axial N$_2$), and finally a proton shift to reach the energetically most favorable intermediate, the Schrock intermediate of (2H,1+)$_1$.

The reduction of the intermediates in (2H,1+)$_1$ leads to a subnetwork in which not the Schrock intermediate but intermediate d) is the most stable intermediate.
This intermediate can be reached through several different cascades of transformations, which however all contain at least one that is comparatively high in energy. 
It can be seen in Fig.~\ref{fig:network} that once an intermediate of (2H,0)$_1$ other than the Schrock intermediate is protonated, no rearrangement reaction within (3H,1+)$_1$ was found which leads to the Schrock intermediate.
This also suggests that the Schrock intermediate in (3H,1+)$_1$, which is relatively high in energy, does not easily transform into a more stable intermediate of the same subnetwork.
Likewise, (3H,1+)$_1$ and (3H,0)$_1$ (not shown in Fig.~\ref{fig:network}) can be considered relevant for the process of degradation of this catalyst.

After reduction of the Schrock intermediate of (3H,1+)$_1$, NH$_3$ dissociates and the [Mo]-N complex is formed.
Similar to the first half, the protonation of the [Mo]-N complex can lead to two different intermediates: the Schrock intermediate and the [Mo]-N complex with a proton at one of the amido groups.
These two structures could give rise to two different reaction paths.
Furthermore, two other subnetworks appear to be particularly prone to initiating degradation: (3H,1+)$_2$ and (3H,0)$_2$.
In both subnetworks there are intermediates that are more stable than the Schrock intermediate, which can be reached via low-energy transition states.
In (3H,1+)$_2$, it is the shift of one of the three protons from one of the axial nitrogen atoms to an amido group (see structures e) and f)).
Either of these structures can undergo an additional transformation where the proton bound to the amido group shifts to the molybdenum center.
After reduction, this structure forms intermediate g)---the most stable conformation of (3H,0)$_2$.
Likewise, the Schrock intermediate of (3H,0)$_2$ can undergo a proton shift to form intermediate g).
As mentioned earlier, the dissociation of NH$_3$ from [Mo]-NH$_3$ is highly endothermic \cite{LeGuennic2005,Schenk2008,Schenk2009-2}, and therefore, 
the exchange of NH$_3$ and N$_2$ via a six-coordinated complex is likely to occur.
Therefore, the intermediate g) in (3H,0)$_2$ can be considered particularly relevant to understanding the low turnover number of the catalyst.

It should be evident from the presentation above that our automated visualization strategy generates a presentation of chemical reaction networks 
that directly unveils its essence to the reader.
Thereby, even complex reaction mechanisms involving many side reactions become lucid and, hence, will be comprehensible.

\subsection{Alternative Pathways to the Chatt--Schrock Cycle}
\begin{figure}[H]
\begin{center}
\includegraphics[width=\textwidth]{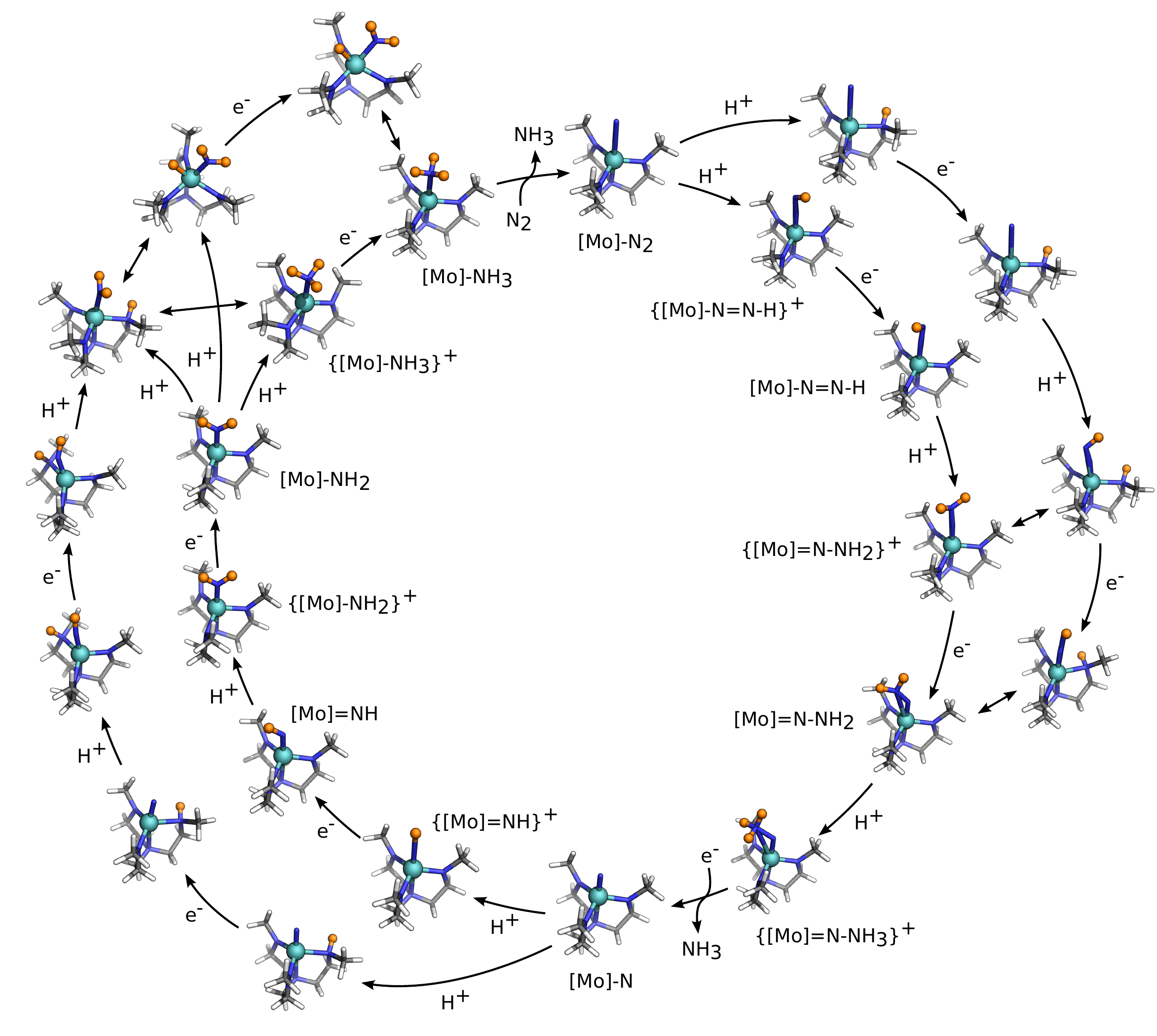}
\caption{
Alternative pathways (outer circle) to the Chatt--Schrock cycle (inner circle).
Proton shifts are indicated by double-headed arrows.
Element color code: gray, C; blue, N; turquoise, Mo; white, H; orange, H added to reactive sites.
\label{fig:ex_chatt_schrock}
}
\end{center}
\end{figure}

By applying the energy-cutoff rule together with the cutoff $E_\text{C}$ = 25 kcal/mol, many intermediates in the reaction network can be removed.
The resulting reaction network allows for the identification of reaction pathways, other than the Chatt--Schrock cycle, that are likely to occur at ambient conditions.
These pathways are shown in Fig.~\ref{fig:ex_chatt_schrock}.

It can be seen that multiple pathways next to the Chatt--Schrock cycle are indeed possible.
For example, two pathways running parallel to the Chatt--Schrock cycle can be identified.
It is important to note that pathways which do not form a cycle, and thus lead to the degradation of the catalyst, are not shown, but will be investigated in 
future studies.

\section{Conclusions}

In this work, a heuristics-guided protocol for the automatic exploration of chemical reaction spaces is presented.
Our heuristics-guided search for chemical species is based on an intuitive chemical construction principle.
A target species (e.g., a catalyst scaffold) reacts with a reactive species (e.g., a radical or a charged particle) to yield an intermediate.
A collection of all intermediates is arranged in a reaction network.
The resulting chemical reaction network can be pruned by defining an energy cutoff, which allows for the exclusion of those 
intermediates which are inaccessible under a range of reasonable physical reaction conditions and on the time scale of interest.

Our protocol for finding vertices in the reaction network exploits conceptual electronic-structure theory to apply heuristic rules for the search of 
potential products in complex reaction mechanisms. The heuristic rules guide the construction of reactive complexes that
are high-energy structures of supermolecules built from reactants from which reaction products (intermediates) are produced upon structure optimization.
The structures of these intermediates enter an emerging reaction network, in which elementary reactions can be  
identified in an automated way.

There are several advantages of our heuristics-guided exploration protocol in practice.
Many calculations of the steps involved --- generation and optimization of reactive-complex structures, 
elementary-reaction proposition, tran\-si\-tion-state search, and network pruning ---
can be carried out independently, which facilitated a trivial parallelization of the whole procedure.
The setup of structures that resemble reactive complexes based on concepts from electronic-structure theory
make the approach applicable to arbitrary reactions and is not limited to any sort of molecule.
Our approach is also efficient for iterative or nested explorations in which the heuristic search restarts from higher-generation structures.

We applied our heuristics-guided exploration protocol to the Chatt--Schrock nitrogen-fixation cycle. Its competing reaction paths
were not studied in sufficient detail. We explored a vast number of
possible elementary reactions that describe protonation, proton-rearrangement, and reduction steps. The resulting network turned
out to be highly complex and alternative routes that still sustain the catalytic cycle emerge. 
The application of an automated visualization strategy by which thermodynamic and kinetic network properties 
was crucial to facilitate the interpretation of such complex reaction mechanisms.
In future work, we will consider the possible degradation reactions that may deactivate the catalyst and that may eventually
explain its low turnover number.

\section*{Appendix: Computational Methodology}
\label{sec:raw_data}

Restricted and unrestricted density-functional-theory calculations were carried out depending on the lowest spin multiplicity of a given intermediate.
For this, BP86/def2-SV(P) \cite{Becke1988,Perdew1986,Weigend2005} structure optimizations of reactive complexes were performed with the program package \textsc{Turbomole} \cite{Ahlrichs1989} (version 6.4.0) including the resolution-of-the-identity density-fitting technique.
Single-point calculations were considered to be converged when the total electronic-energy difference between two iteration steps was less then 10$^{-7}$ Hartree.
Structure optimizations were considered converged when the norm of the electronic-energy gradient with respect to the nuclear coordinates dropped below 10$^{-4}$ Hartree/Bohr.
If a structure optimization failed because a self-consistent-field calculation did not converge, the damping parameters had been changed automatically and the optimization was restarted. 
In those cases where a structure optimization did not converge within 1200 iterations, the corresponding data was saved and the structure was manually inspected to decide whether it should be part of the chemical reaction network or not.
9607 structure optimizations were carried out in total.

Constrained BP86/def2-SV(P) optimizations were performed wih \textsc{Gaussian} \cite{g09} (version 09, revision C.1) to obtain reasonable starting structures of transition states, which were refined with \textsc{Turbomole}'s trust-radius-image-based eigenvector-following optimization choosing a trust-radius of 0.2 {\AA}.
The eigenmode to follow was obtained from a Mode-Tracking calculation \cite{Reiher2003, Bergeler2015}. 
From the converged transition states, intrinsic reaction paths were calculated with \textsc{Gaussian} to determine whether a desired transition state was found.
We employed the default convergence criteria for all \textsc{Gaussian} calculations.
If the constrained optimization scan with a subsequent eigenvector-following calculation did not converge to the desired transition state, the freezing-string method as implemented in \textsc{Q-Chem} (version 4.0.1) \cite{Shao2006} was employed with subsequent EVF as described above.
We identified 2318 elementary reactions for which transition states were optimized.

To shed more light on the success rate of identifying transition states for these reactions by our automated search and therefore of verifying the
assumption that two intermediates are truly connected by an elementary reaction, we may add some additional details.
1082 potential elementary reactions were automatically identified for the first and 1236 for the second half of the Chatt--Schrock cycle.
The transition-state search was then conducted with three different strategies yielding a total of 6954 transition-state searches. This
number, however, is only an upper bound as a search was stopped once one of these strategies was successful. 
In the first half of the cycle, our automated protocol identified 329 out of the 1082 potential elementary reactions by optimizing the transition state.
In the second half, it identified 613 out of the 1236 potential elementary reactions.
For some steps, for which our implementation was not able to find a transition state, we verified by manual inspection that 
a transition state is not likely to exist or to be of sufficiently low energy. Hence, our algorithm produces more
potential pairs that could be connected by an elementary reaction than there are. Of course, this number is determined by
the structural similarity measure that we employ to relate the two structures. Obviously, our RMSD criterion produces many
false positive results. However, this is actually desired as one cannot be certain to have found all relevant vertices and edges
of a reaction network so that all criteria and measures should be set and selected in a conservative and therefore not too restrictive way.

The calculation of the ELF and of the electrostatic potential were performed with {\sc Molden} 5.4 \cite{molden}.
In the color range of the electrostatic potential mapped onto the ELF isosurface, 
the most positive charge was omitted as otherwise the color differences between all other atoms would have been very small.
The presentation of the data in Fig.\ \ref{elf} was generated with {\sc Jmol} 14.0.7 \cite{jmol} from a cube file produced with {\sc Molden}. 

The visualization of all (sub)networks was automated in the programming language Python. For their representation, the Python software package 
NetworkX \cite{Hagberg2008} was applied.

\section*{Acknowledgments}
This work was financially supported by ETH Z\"urich (Grant ETH-20 15-1) and by the Schweizerischer Nationalfonds (SNF project 200020\_156598). 
MB and GNS gratefully acknowledge support by PhD fellowships of the Fonds der Chemischen Industrie.

\section*{Supporting Information Available}
Additional figures of the reaction network with further details on species and structures are collected in the supporting information.
This information is available free of charge via the Internet at http://pubs.acs.org/.


\providecommand{\url}[1]{\texttt{#1}}
\providecommand{\urlprefix}{}
\providecommand{\foreignlanguage}[2]{#2}
\providecommand{\Capitalize}[1]{\uppercase{#1}}
\providecommand{\capitalize}[1]{\expandafter\Capitalize#1}
\providecommand{\bibliographycite}[1]{\cite{#1}}
\providecommand{\bbland}{and}
\providecommand{\bblchap}{chap.}
\providecommand{\bblchapter}{chapter}
\providecommand{\bbletal}{et~al.}
\providecommand{\bbleditors}{editors}
\providecommand{\bbleds}{eds.}
\providecommand{\bbleditor}{editor}
\providecommand{\bbled}{ed.}
\providecommand{\bbledition}{edition}
\providecommand{\bbledn}{ed.}
\providecommand{\bbleidp}{page}
\providecommand{\bbleidpp}{pages}
\providecommand{\bblerratum}{erratum}
\providecommand{\bblin}{in}
\providecommand{\bblmthesis}{Master's thesis}
\providecommand{\bblno}{no.}
\providecommand{\bblnumber}{number}
\providecommand{\bblof}{of}
\providecommand{\bblpage}{page}
\providecommand{\bblpages}{pages}
\providecommand{\bblp}{p}
\providecommand{\bblphdthesis}{Ph.D. thesis}
\providecommand{\bblpp}{pp}
\providecommand{\bbltechrep}{Tech. Rep.}
\providecommand{\bbltechreport}{Technical Report}
\providecommand{\bblvolume}{volume}
\providecommand{\bblvol}{Vol.}
\providecommand{\bbljan}{January}
\providecommand{\bblfeb}{February}
\providecommand{\bblmar}{March}
\providecommand{\bblapr}{April}
\providecommand{\bblmay}{May}
\providecommand{\bbljun}{June}
\providecommand{\bbljul}{July}
\providecommand{\bblaug}{August}
\providecommand{\bblsep}{September}
\providecommand{\bbloct}{October}
\providecommand{\bblnov}{November}
\providecommand{\bbldec}{December}
\providecommand{\bblfirst}{First}
\providecommand{\bblfirsto}{1st}
\providecommand{\bblsecond}{Second}
\providecommand{\bblsecondo}{2nd}
\providecommand{\bblthird}{Third}
\providecommand{\bblthirdo}{3rd}
\providecommand{\bblfourth}{Fourth}
\providecommand{\bblfourtho}{4th}
\providecommand{\bblfifth}{Fifth}
\providecommand{\bblfiftho}{5th}
\providecommand{\bblst}{st}
\providecommand{\bblnd}{nd}
\providecommand{\bblrd}{rd}
\providecommand{\bblth}{th}

\end{document}